\documentclass[lettersize,journal,10pt]{IEEEtran}

\usepackage[utf8]{inputenc}
\usepackage[english]{babel}
\usepackage{booktabs}
\usepackage{multirow}
\usepackage{amssymb}
\usepackage{algorithm,algorithmic}
\usepackage{cite}
\usepackage{float}
\usepackage[final]{graphicx}
\usepackage{xcolor}
\usepackage{epsfig}
\usepackage{tabularx}
\usepackage{makecell}
\usepackage{subcaption}
\usepackage{pifont}
\definecolor{mplgreen}{HTML}{2E7D32}
\definecolor{mplred}{HTML}{C62828}

\usepackage[cmex10]{amsmath}
\usepackage[cmintegrals]{newtxmath}

\usepackage{balance}

\usepackage{amsmath}
\usepackage{siunitx}

\begin{document}
\pagenumbering{gobble}
\bstctlcite{IEEEexample:BSTcontrol}

\title{
On the Value of Base Station Motion Knowledge for Goal-Oriented Remote Monitoring with Energy-Harvesting Sensors
}

\author{\IEEEauthorblockN{Sehani~Siriwardana, Jean~M.~S.~Sant'Ana, Richard~Demo~Souza, Abolfazl~Zakeri, Onel L. A. López}
\\
\IEEEauthorblockA{\textit{Centre for Wireless Communications, University of Oulu}, Oulu, Finland\\
\textit{Electrical \& Electronics Engineering Department, Federal University of Santa Catarina}, Florianópolis, Brazil\\
stharindra@gmail.com,
\{jean.desouzasantana, abolfazl.zakeri, onel.alcarazlopez\}@oulu.fi,
richard.demo@ufsc.br}}

\maketitle

\begin{abstract}
This paper investigates goal-oriented remote monitoring of an unobservable Markov source using energy-harvesting sensors that communicate with a mobile receiver, such as a Low Earth Orbit (LEO) satellite or Unmanned Aerial Vehicle (UAV). Unlike conventional systems that assume stationary base stations, the proposed framework explicitly accounts for receiver mobility, which induces time-varying channel characteristics modeled as a finite-state Markov process. The remote monitoring problem is formulated as a partially observable Markov decision process (POMDP), which is transformed into a tractable belief-state MDP and solved using relative value iteration to obtain optimal sampling and transmission policies. Two estimation strategies are considered: Maximum Likelihood (ML) and Minimum Mean Distortion (MMD). Numerical results demonstrate that incorporating receiver mobility and channel state information into the optimization reduces the average distortion by 10\% to 42\% compared to baseline policies and constant-channel assumptions, highlighting the importance of base station motion knowledge for effective goal-oriented communication.
\end{abstract}

\section{Introduction}

Internet of Things (IoT) applications are presented in several areas of the world. Some consist of remote areas, beyond cellular network coverage, energy grid availability, or areas with difficult maintenance access, which may require on-site energy generation methods, such as energy harvesting~\cite{liu2021recent}. In such cases, energy consumption reduction is crucial, motivating the research on efficient sampling and data transmission, so that only the information necessary to achieve a specific objective is collected and transmitted, in which is referred to as goal-oriented remote monitoring~\cite{getu2024survey}.

In cases with limited cellular coverage, other solutions can be used to serve as base stations. Some examples includes Low-Earth Orbit (LEO) satellites~\cite{he2024non}, Unmanned Aerial Vehicles (UAVs)~\cite{Ijemaru:electronics:2021, Wei:IoTJ:2022} and maritime vessels~\cite{Wei:IoTJ:2021}, in which can be used as an enabler for IoT connectivity in remote regions. All of the aforementioned examples consist of mobile base stations. In those cases, the performance of the transmission between devices and the monitor at the base station is directly connected to their movement and position. Therefore, this paper investigates goal-oriented remote monitoring of an unobservable source using sensors deployed in remote regions that transmit to mobile monitors.

In~\cite{pappas:2021:ICAS}, a goal-oriented real-time remote monitoring framework is presented for autonomous systems. Its system model consists of an information source, a sampler, a transmitter, and a stationary receiver.
Zakeri et al. ~\cite{zakeri:TWC:2024} considered a system wherein a transmitter sends information from multiple sources to a monitor through a relay. They formulated a constrained Markov decision process (CMDP), which is then transformed into a standard Markov decision process (MDP) and solved using a structure-aware relative value iteration algorithm (RVIA). Later, the work in~\cite{zakeri2024goal} presented a goal-oriented remote tracking system with energy constraints, where the system comprises an information source, a sensor, and a monitor. Therein, the information source and sensor are deployed in a remote area, and the sensor is powered by energy harvesting. The information source is modeled as a Markov chain with several states, and the overall system is formulated as a partially observable MDP (POMDP). Then, this POMDP is transformed into an MDP and solved using a RVIA.

The aforementioned works considered a \textit{stationary} monitor, where the channel state is constant. This may hold true in standard cellular communication or in GEO satellite systems, where the satellite orbit follows Earth's rotation and covers the same area of the globe. However, in scenarios like LEO satellite systems or UAV communication, the monitor appears to move from the transmitter’s point of view, generating a time-varying channel. To better model this scenario, the motion of the base stations should be incorporated. In previous research~\cite{lutz:1996:Wiley,aboderin:2015:modeling,rougerie:2016:EuCAP,akinniyi:2017:Parallel,lee:2019:Access}, satellite motion is modeled as a binary Markov chain with two states: good and bad. The good state represents a line-of-sight (LoS) condition, whereas the bad state corresponds to a non-line-of-sight (NLoS) condition. E. Lutz~\cite{lutz:1996:Wiley} further proposes a four-state Markov chain model in which they model satellite diversity and their channel correlation. In another study~\cite{fontan:2001:statistical}, a three-state Markov chain was introduced, where the states are defined based on the variation of the channel quality as the satellite moves. Specifically, state 1 represents a clear LoS condition between the satellite and the transmitter, state 2 corresponds to moderate shadowing, and state 3 denotes a complete absence of LoS. Furthermore, another study~\cite{hui:2008:new} proposed a five-state Markov chain model, with the states defined according to the motion dynamics of the satellite.

In this work, we propose a goal-oriented remote monitoring system for satellite communications that accounts for the base station mobility inherent in LEO and UAV systems. To account that, we extend the framework proposed in~\cite{zakeri2024goal}. The general operation of the system is formulated as a POMDP, which is then transformed into an MDP. The resulting MDP is solved to obtain the optimal sensor policies for sampling and transmission. We analyze the impact of optimal policies in terms of distortion based on a given source application. Our results show that the optimized average distortion can be reduced from 10\% to 42\% when considering the base station motion behavior in the optimization.

\begin{figure*}[!th]
    \centering
    \includegraphics[width=0.9\linewidth]{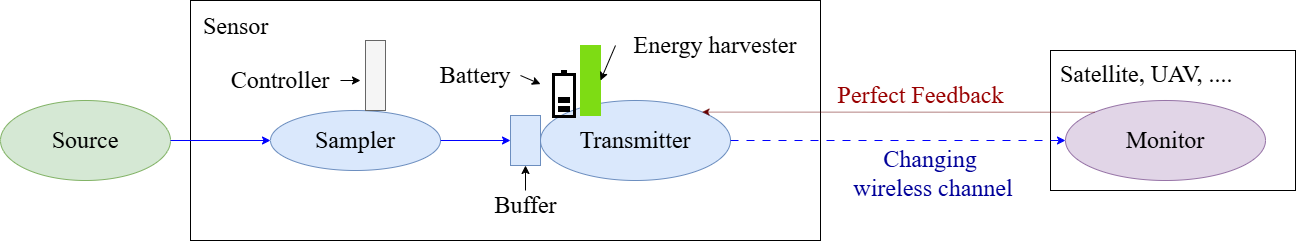}
    \caption{System model illustrating the sensor components (sampler, buffer, transmitter, controller), energy harvesting module, and mobile monitor connected through a time-varying wireless channel.}
    \label{fig:system_model_new}
    \vspace{-0.5cm}
\end{figure*}

\section{System Model}
\label{sec:System_Model}

We consider a system model similar to the one in \cite{zakeri2024goal} as illustrated in Figure~\ref{fig:system_model_new}.
The model consists of an information source, a sensor, and a monitor. The sensor comprises a sampler, a controller, a buffer, and a transmitter. The sensor operation is powered by a battery that is continuously charged using an energy harvesting method. The sampler first samples the information from the information source, stores the sampled information in the buffer, and then the transmitter sends the buffered information to the moving monitor through a wireless channel. Next, within the sensor, the controller regulates the sampling and transmission processes by issuing commands to the sampler and the transmitter. It is assumed that the controller observes the battery level, the information in the buffer, and the success or failure of each transmission event. Since the monitor is in motion, the wireless channel varies with its position in the sky relative to the sensor; consequently, the reception success rate (RSR) also varies.

The information source is modeled as a Markov source with $M$ states. The entire system operates in discrete time with time slots $t \in \{0, 1, 2, \dots\}$.
Then, the information state at the source at each time slot $t$ is denoted by $X_t$, where $X_t \in S_X=\{1, 2, \dots, M\}$, and $S_X$ is the state space. Then, the information source state transition matrix is $P = \big[p_{ij}\big]$, where $p_{ij}$ represents the probability of transitioning from the state $i$ to the state $j$. Next, the information state in the buffer is indicated by $\tilde{X}_t$, while the most recent information received at the monitor is indicated by $\bar{X}_t$.

The channel is modeled as a Markov source with $K$ states to mimic the monitor motion where the RSR varies over time. The motion transition matrix is denoted by  $\mathcal{Q} = [p_{kl}]$, where $p_{kl}$ represents the probability of transitioning from state $k$ to the state $l$, and the state of the monitor at time $t$ is denoted by $Q_t \in \{1, 2, \dots, K\}$. Finally, we have the RSR at time $t$ represented as $q_s(Q_t)$, where $q_s$ is a vector of length $K$ that contains RSR for each respective state. It is important to note that when $K=1$, there is only one monitor state and the RSR is constant, and the model converges to the one in~\cite{zakeri2024goal}.

Then, the controller commanding process for each sampling and transmission at each time slot is formulated as a binary decision model, represented by $\alpha_t$ and $\beta_t$, respectively. Precisely, $\alpha_t = 1$ defines that the sampler collects a sample from the source, while $\alpha_t = 0$ indicates that no sampling occurs. Similarly, $\beta_t = 1$ denotes that the transmitter sends the currently stored sample $\Tilde{X}_t$ from the buffer to the monitor, while $\beta_t = 0$ indicates that no transmission has occurred. Thus, it forms a combination of 4 possible actions in each time slot $t$, as shown in Table~\ref{tab:actions}.

\begin{table}[b]
\caption{Controller actions.}
\centering
\begin{tabular}{|c|c|c|c|}
\hline
\textbf{$\alpha_t$ and $\beta_t$} & \textbf{Action} $a$ & cost $c(a)$ & \textbf{Controller action} \\
\hline
 $\alpha_t = 0 $, $\beta_t = 0 $ & 0 & 0 &No sampling and no transmitting  \\
 $\alpha_t = 0 $, $\beta_t = 1 $ & 1 & $\tau$ &No sampling and just transmitting \\
 $\alpha_t = 1 $, $\beta_t = 0 $ & 2 & $\kappa$ &Just sampling and no transmitting \\
 $\alpha_t = 1 $, $\beta_t = 1 $ & 3 & $\tau+\kappa$ &Both sampling and transmitting \\
\hline
\end{tabular}
\label{tab:actions}
\end{table}

We assume energy is consumed only during two operations: (i) sampling the source information state, and (ii) transmitting the buffered information state to the monitor. The corresponding energy costs are denoted by $ \kappa $ for sampling and $ \tau $ for transmission. Moreover, energy must be harvested from the environment and it is stored in a battery with discrete energy levels, where the maximum capacity is $ E $. The energy arrival process, denoted by $ u_t $, is modeled as a Bernoulli process following ~\cite{privault:2008:stochastic}, with an energy arrival rate $\mu$, such that $ \mathbb{P}\{u_t = 1\} = \mu $. Moreover, the battery state at time slot $ t $, denoted by $ e_t \in \{0, \dots, E\} $, evolves according to
\begin{equation}
    e_{t+1}=\min \left\{e_t+u_t-\left(\kappa \alpha_t+\tau \beta_t\right), E\right\}.
\label{eq:battery_energy_level_update}
\end{equation}
At each time slot $t$, executing an action requires sufficient energy available in the battery, which can be expressed as
\begin{equation}
    e_t-\alpha_t \kappa-\beta_t \tau \geq 0, \forall t.
\label{eq:sufficient_energy_level_checking}
\end{equation}

We define a cost function $d_t$, which incorporates an estimate of the information source state, along with the information state at the source during the same time slot $t$, which is expressed as
\begin{equation}
d_t \triangleq f(X_t, \hat{X}_t) ,
\label{eq:distortion_function}
\end{equation}
where $\hat{X}_t$ denotes the monitor estimation of the source information state $X_t$. Then, to obtain the estimation of $X_t$, two approaches are considered: Maximum Likelihood (ML)~\cite{eliason1:993:Sage} and Minimum Mean Distortion (MMD) estimations. Moreover, the cost function reflects the cost incurred at each time slot when estimation is required whenever the source information is not sampled, transmitted, and received successfully by the monitor within that time slot. For optimized system performance, the cost must be minimized, as successful sampling and transmission within the same time slot are required, subject to the constraint in \eqref{eq:sufficient_energy_level_checking}.
The system problem can be formulated~as
\begin{equation}
\begin{array}{ll}
\operatorname{minimize} & \limsup _{T \rightarrow \infty} \frac{1}{T} \sum_{t=1}^T \mathbb{E}\{d_t\} \\
\text {subject to} & e_t-\alpha_t \kappa-\beta_t \tau \geq 0, \forall t,
\end{array}
\label{eq:problem formulation}
\end{equation}
where $\sum_{t=1}^T \mathbb{E}\{\cdot\}$ represents the sum of the expected cost obtained with respect to the randomness of the system, which includes the source, the battery charging and the channel. By solving this problem, we get the optimal action values $\alpha_t$ and $\beta_t$ that should be taken at each time slot.

\section{Problem Solution}
\label{sec:new_Transformation}

Since the controller does not directly access the state of the information source, there is only partial observability. To account for this, the system dynamics are modeled as a POMDP, which is subsequently transformed into an equivalent Markov Decision Process (MDP) and solved using RVIA, which facilitates the determination of optimal control policies.

The elements of the POMDP are defined next.

\textbf{State Space ($S$)}: In each time slot $t$, the system state is represented by 
$s_t = \big(e_t, X_t, \tilde{X}_t, \bar{X}_t, \theta_t, \delta_t, Q_t\big)$ and $S$ is an infinite set, where $\theta_t$ is the age of information (AoI) at the transmitter (age of the last sampled state at the buffer) and $\delta_t$ is the AoI at the monitor (age of the last received sample at the monitor).

\textbf{Action Space ($\mathcal{A}_s$)}: Let $\mathcal{A}_s$ be the action space that consists of all admissible actions that the controller can take, i.e., actions satisfying the energy constraint, when the system is at state $s$. The action selected at time slot $t$ is (generally) represented as $a_t \in \{0,1,2,3\}$, where
$a_t = 0$ indicates that the sampler and transmitter stay idle,
$a_t = 1$ indicates that the transmitter re-transmits the sample
in the buffer, $a_t = 2$ indicates that the sampler takes a
new sample, and $a_t = 3$ indicates that the sampler takes a
new sample and the transmitter transmits that sample. The actions are determined by a policy $\pi$, which maps $S$ to $\mathcal{A}_s$.

\textbf{Observation Space ($\Omega$)}: The observation space $\Omega$ represents the set of all observations available to the controller. The observation $o_t$ at time slot $t$ is given by  $o_t = (e_t, \tilde{X}_t, \bar{X}_t, \theta_t, \delta_t, Q_t)$.

\textbf{State Transition Probabilities ($P_{\text{POMDP}}$)}:  
The state transition probabilities characterize the chance of moving from the current state  
$s_t = \big(e_t, X_t, \tilde{X}_t, \bar{X}_t, \theta_t, \delta_t, Q_t\big)$  
to the next state  
$s_{t+1} = \big(e_{t+1}, X_{t+1}, \tilde{X}_{t+1}, \bar{X}_{t+1}, \theta_{t+1}, \delta_{t+1}, Q_{t+1}\big),$ 
conditioned on the action $a_t$ selected at time slot $t$. This probability is denoted by $\mathbb{P}\{s_{t+1}\,|\,s_t,a_t\}$. Since the system processes including the dynamics of the information source, monitor motion, AoI evolution, battery states, and source estimation are assumed to evolve independently, the overall transition probability can be represented as the product of the transition probabilities of these individual components as

\begin{equation}
\begin{aligned}
&\mathbb{P}\{s_{t+1} \mid s_t, a_t\} =\mathbb{P}\left\{\theta_{t+1} \mid \theta_t, a_t\right\} \\
& ~~ \times \mathbb{P}\left\{\delta_{t+1} \mid \theta_t, \delta_t, a_t\right\} \, \mathbb{P}\left\{e_{t+1} \mid e_t, a_t\right\} \\
& ~~ \times \mathbb{P}\left\{\tilde{X}_{t+1} \mid X_t, \tilde{X}_t, a_t\right\} \, \mathbb{P}\left\{X_{t+1} \mid X_t\right\} \\ 
& ~~ \times\mathbb{P}\left\{Q_{t+1} \mid Q_t\right\}  \, \mathbb{P}\left\{\bar{X}_{t+1} \mid X_t, \tilde{X}_t, \bar{X}_t, a_t\right\},
\end{aligned}
\label{eq:statetransition_probabilitiesnew}
\end{equation}
where
\begin{equation}
\mathbb{P}\{\theta_{t+1}\mid \theta_t,a_t\}
=
\delta_{\theta_{t+1}}
\,\mathbf{1}_{\{a_t\in\{2,3\}\}}
+
(\theta_t+1)\mathbf{1}_{\{a_t\in\{0,1\}\}
},
\end{equation}
\begin{equation}
\begin{aligned}
&\mathbb{P}\{\delta_{t+1} \mid \theta_t,\delta_t,Q_t,a_t\}= \\&~~
\begin{cases}
q(Q_t), & \text{if } a_t=3,\; \delta_{t+1}=1,\\
q(Q_t), & \text{if } a_t=1,\; \delta_{t+1}=\theta_t+1,\\
\bar q(Q_t), & \text{if } a_t\in\{1,3\},\; \delta_{t+1}=\delta_t+1,\\
1, & \text{if } a_t\in\{0,2\},\; \delta_{t+1}=\delta_t+1,\\
0, & \text{otherwise.}
\end{cases}
\end{aligned}
\end{equation}

\begin{equation}
\begin{aligned}
&\mathbb{P}\left\{\tilde{X}_{t+1} \mid X_t, \tilde{X}_t, a_t\right\}=  
\begin{cases}
1, & \text{if } a_t \in \{2,3\}, \tilde{X}_{t+1} = X_t, \\
1, & \text{if } a_t \in \{0,1\}, \tilde{X}_{t+1} = \tilde{X}_t, \\
0, & \text{otherwise,}
\end{cases}    
\end{aligned}
\end{equation}
\begin{equation}
\mathbb{P}\left\{X_{t+1} \mid X_t\right\} = p_{X_t, X_{t+1}},
\end{equation}
\begin{equation}
\mathbb{P}\left\{Q_{t+1} \mid Q_t\right\} = p_{Q_t, Q_{t+1}},
\end{equation}
\begin{equation}
\begin{aligned}
&\mathbb{P}\left\{\bar{X}_{t+1} \mid X_t,\tilde{X_t}, \bar{X_t}, Q_t, a_t\right\}=\\ ~~&
\begin{cases}
q(Q_t), & \text{if } a_t = 3, \bar{X}_{t+1} = X_t\\
q(Q_t), & \text{if } a_t = 1, \bar{X}_{t+1} = \tilde{X}_t \\
\bar{q}(Q_t), & \text{if } a_t \in \{1,3\}, \bar{X}_{t+1} = \bar{X}_t \\
1, & \text{if } a_t \in \{0,2\}, \bar{X}_{t+1} = \bar{X}_t \\
0, & \text{otherwise,}
\end{cases}
\end{aligned}
\end{equation}
\begin{equation}
\mathbb{P}\{e_{t+1} \mid e_t, a_t\}=
\begin{cases}
\mu, & e_{t+1}=\min\{e_t+1,E\}-c(a_t),\\
\bar\mu, & e_{t+1}=e_t-c(a_t),\\
0, & \text{otherwise,}
\end{cases}
\end{equation}
where $\bar{q}(Q_t) = 1 - q(Q_t)$, $\bar{\mu} = 1 - \mu$, $Q_t \in {1, 2, 3, \dots, K}$ and $c(a)$ is the energy cost in respective to action $a$. 

\textbf{Observation Function ($O$)}: The observation function specifies the likelihood of observing $o_t$ at time $t$, conditioned on the current state $s$ and the action $a_{t-1}$ executed in the previous time slot. It can be written as
\begin{equation}
O(o_t \mid s_t, a_{t-1}) = \mathbb{P}\{ o_t \mid s_t, a_{t-1} \}.
\end{equation}

\textbf{Cost Function ($C$)}: Following~\cite{zakeri2024goal}, the cost function $C(s_t)$, representing the cost at time slot $t$ is given by
\begin{equation}
C(s_t) = f(X_t, \hat{X}_t),
\end{equation}
where the estimate $\hat{X}_t$ depends on the most recently received sample on the monitor, $\bar{X}_t$, and the AoI on the monitor, $\delta_t$.

\textbf{Initial Belief State ($b^0$)}: The initial belief state $b^{0}$ defines the probability distribution over all possible system states at the start of the process.

\subsection{MDP Reformulation}

As discussed in~\cite{sigaud:Wiley:2013}, the solution of a POMDP can be approached by reformulating it as a MDP through the introduction of a complete information state, $\mathcal{I}_t$. The complete information state $\mathcal{I}_t$ is composed of three elements: the initial probability distribution over the state space, the sequence of observations up to time $t$, i.e., $\{o_0, o_1, \ldots, o_t\}$, and the set of actions executed by the controller until time $t-1$, i.e., $\{a_0, \ldots, a_{t-1}\}$. Based on this representation, a belief state $b^{i}_t$ is defined at each time slot $t$, which describes the probability that the system is in state $i$, conditioned on the complete information state $I_t$. The belief state is
\begin{equation}
b^{i}_t \triangleq \mathbb{P}\left\{X_t=i \mid \mathcal{I}_t\right\}, i=1, \ldots, M.
\end{equation}
According to Proposition~1 in~\cite{zakeri2024goal}, given the current belief state $b^{i}_t$ and the chosen action $a_t$ at time slot $t$, the subsequent belief state $b^{i}_{t+1}$ is obtained by
\begin{equation}
\begin{aligned}
b^{i}_{t+1}=
\begin{cases}
b^{i}_t p_{i i}+\sum_{j\,j \neq i} b^{j}_t p_{j i} & \text{if\;} a_t \in\{0,1\}, \\
p_{\tilde{X}_{t+1} i} & \text{if\;} a_t \in\{2,3\},
\end{cases}
\end{aligned}
\label{eq:beliefupdatenew}
\end{equation}
where $i = 1, \dots, M$ and $j = 1, \dots, M$. 

Then, in the transformation from the POMDP into an equivalent MDP, the belief state is integrated into the original POMDP structure, and the components of the resulting MDP are defined as described next.

\textbf{State Space ($Z$)}: The MDP state space is denoted by $Z$, and the state at a given time slot $t$ is represented by $z_t$ as
\begin{equation} 
z_t \triangleq \left(e_t, \left\{b_{i}(t)\right\}_{i=1, \ldots, M}, \tilde{X}_t, \bar{X}_t, \theta_t, \delta_t, Q_t\right), \label{eqn:state_space_z}
\end{equation}
where $\left\{b^{i}_t\right\}_{i=1, \ldots, M}$ denotes the belief distribution over the source states. Moreover, since the time horizon $T$ is infinite and the belief state $\left\{b^{i}_t\right\}_{i=1, \ldots, M}$ is continuous, the state space $Z$ is inherently infinite.

\textbf{Action Space ($A$)}: The action space of the belief MDP remains identical to that of the POMDP model.

\textbf{State Transition Probabilities ($P_{\text{MDP}}$)}:  
The transition probability from the current state $z_t$ to state $z_{t+1}$ is given by
\begin{equation}
\begin{aligned}
\mathbb{P}\{z_{t+1} \mid z_t, a_t\} &= \sum_{X_t \in \mathcal{X}} 
\mathbb{P}\{z_{t+1} \mid z_t, a_t, X_t\} \mathbb{P}\{X_t \mid z_t, a_t\}.
\label{eq:statetransitionprobabilitynew}
\end{aligned}
\end{equation}
Here, the probability term $\mathbb{P}\{z_{t+1} \mid z_t, a_t, X_t\}$ can be obtained using~\eqref{eq:statetransition_probabilitiesnew} together with~\eqref{eq:beliefupdatenew}.  
In addition, the observation probability is expressed as  
$\mathbb{P}\{X_t \mid z_t, a_t\} = b_{i}(t)$, for $i = 1, \ldots, M$.

\textbf{Cost Function ($r$)}:  
The cost function that evaluates the expected cost at time slot $t$, is expressed as
\begin{equation}
C\big(z_t\big) = \sum_i b^{i}_t \, f\big(i, \hat{X}_t\big),
\label{eq:reward function new new MDP}
\end{equation}
where $f\!\left(i, \hat{X}_t\right)$ denotes the cost obtained when the source state $i$ is estimated as $\hat{X}_t$.  
Using these definitions, we focus on solving the belief MDP problem 
\begin{equation}
\pi^* = \arg\min_{\pi \in \Pi}
\left(\limsup_{T \to \infty} \frac{1}{T} \sum_{t=1}^{T} \mathbb{E}\{ C(z_t) \}
\right),
\label{eq:pi_opt}
\end{equation}
where $\pi^*$ is the optimal policy and $\Pi$ is the set of all policies that satisfy the energy constraint~\eqref{eq:sufficient_energy_level_checking}.

The presence of the continuous state variable $\{b_i(t)\}$ in \eqref{eqn:state_space_z}
makes finding an optimal policy to the above problem extremely
challenging since the associated state space $\mathcal{Z}$ is infinite.
Therefore, the state space must be reduced to a finite set. According to Proposition~2 in~\cite{zakeri2024goal}, the belief at time slot $t$, conditioned on observing $\tilde{X}_t$ and $\theta_t$, is given by
\begin{equation}
b^{i}_t = p_{\tilde{X}_ti}^{\theta_t}, \quad i = 1, \ldots, M,
\end{equation}
where $p_{ij}^{\theta_t}$ represents the $(i,j)$-th entry of the matrix $P^{\theta_t}$.
Then, as $\theta_t$ increases, the matrix $P^{\theta_t}$ converges to a steady-state, and consequently, the belief state $b_{i}(t)$ approaches the steady-state probabilities for all $\tilde{X}_t$ and $i \in S_X$. Similarly, since the computation of $\hat{X}_t$ depends on $P^{\delta_t}$, the range of $\delta_t$ can be bounded by a large constant $\delta_{\max}$. With suitable choices of $\theta_{\max}$ and $\delta_{\max}$, the belief MDP problem can be reformulated as finite-state MDP. As a result, the belief terms $\{b^i_t\}_{i=1}^{M}$ can be removed from~\eqref{eqn:state_space_z}, as the belief is a function of $\theta_t$ and the system state can be redefined as
\begin{equation}
z_t \triangleq \big(e_t, \tilde{X}_t, \bar{X}_t, \theta_t, \delta_t, Q_t\big).
\end{equation}

Similarly to~\cite{Zakeri:2025:TC}, one can show that the  MDP with states in (22) is communicating under which the Bellman optimality equation
\begin{equation}
\begin{aligned}
&U^*(z_t + V^*(z_t)) = \\ &
\min_{a_t \in A} \Big( C(z_t) + \!\sum_{z_{t+1} \in Z}  \mathbb{P} \big\{ z_{t+1} \mid z_t, a_t\big\}  U^*(z_{t+1}) 
  \Big),\; \forall z_t \in Z,
\end{aligned}
\label{eq:bellman_average_reward_2_new}
\end{equation}
exists
and can be solved via RVIA. RVIA
transforms the Bellman’s optimality equation into the following iterative
process, such that for all $\underline{z}\in\underline{\mathcal{Z}}$ and for an iteration index ${ n=1,2,\dots }$, we have
\begin{equation}\label{Eq_RVIal}
\begin{array}{ll}
&
\displaystyle
V^{n+1}(\underline{z}) =
\min_{a \in \mathcal{A}}
\left\{C(\underline{z})+ \underset{\underline{z}^{\prime} \in \underline{\mathcal{Z}}}{\sum} 
\operatorname{Pr}\left\{\underline{z}^{\prime} \mid \underline{z}, a\right\} 
h^n(\underline z')\right\},
\\
&
\displaystyle
h^n(\underline z) = V^n(\underline{z})
- V^n(\underline{z}_{\mathrm{ref}}).
\end{array}
\end{equation}
A detailed description of RVIA can be found in, e.g., Algorithm~1 of~\cite{Zakeri:2025:TC}.

\section{Numerical Evaluation}
\label{sec:Experimental_Evaluation}
In this section we evaluate the system model under a few scenarios. We assume a source with $M=5$ states, where its state transition matrix $P$ is given in~\eqref{eq:transition_matrix}
\begin{equation}
P = 
\begin{bmatrix}
0.1    & 0.6    & 0.2    & 0.05   & 0.05 \\
0.8    & 0.05   & 0.05   & 0.03   & 0.07 \\
0.0125 & 0.0125 & 0.95   & 0.0125 & 0.0125 \\
0.1    & 0.7    & 0.1    & 0.05   & 0.05 \\
0.005  & 0.3    & 0.0475 & 0.0475 & 0.6
\end{bmatrix}.
\label{eq:transition_matrix}
\end{equation}
The distortion matrix $d = [d_{ij}]$ is given in~\eqref{eq:distortion_matrix}, where $d_{ij}$ represents the distortion if $\hat{X}_t = i$ is estimated while the source $X_t = j$
\begin{equation}
d = 
\begin{bmatrix}
0 & 383 & 183 & 70 & 529 \\
611 & 0 & 714 & 419 & 281 \\
55 & 271 & 0 & 88 & 611 \\
327 & 967 & 245 & 0 & 491 \\
818 & 820 & 30 & 258 & 0
\end{bmatrix}.
\label{eq:distortion_matrix}
\end{equation}
In addition, we assume the transmission $\tau$ and sampling $\kappa$ costs are 1, the battery capacity $E=5$ and the AoIs bounds $\theta_{\max} = \delta_{\max} =30$. Finally, we showcase two monitor motion transition matrix $\mathcal{Q}$ and their respective RSR vectors $q_s$ in \eqref{eq:motion1} and \eqref{eq:motion2}
\begin{equation}
\mathcal{Q} = 
\begin{bmatrix}
0.1 & 0.9 \\
1 & 0
\end{bmatrix},
\qquad q_s = 
\begin{bmatrix}
0 & 0.9
\end{bmatrix},
\label{eq:motion1}
\end{equation}
\begin{equation}
\begin{aligned}
\mathcal{Q} &= 
\begin{bmatrix}
0.1 & 0.9 & 0 & 0 & 0 & 0 \\
0 & 0 & 1 & 0 & 0 & 0 \\
0 & 0 & 0 & 1 & 0 & 0 \\
0 & 0 & 0 & 0 & 1 & 0 \\
0 & 0 & 0 & 0 & 0 & 1 \\
1 & 0 & 0 & 0 & 0 & 0
\end{bmatrix}, \\[6pt]
q_s &= 
\begin{bmatrix}
0 & 0.25 & 0.6 & 0.9 & 0.6 & 0.25
\end{bmatrix}.
\end{aligned}
\label{eq:motion2}
\end{equation}
The first represents a very basic scenario where the channel changes between good and bad, with a small chance of staying on the bad state. The second tries to mimic a LEO satellite behavior, where the success probability increases and decreases, like a satellite going from its rise moment, reaching its culmination (maximum elevation) and setting afterwards. Note that, in both cases, we considered a chance (0.1) of the first state to remain in the same state. This represents the case without coverage and no satellite appears in the following slot. It is important to notice that the average RSR (steady-state) in both cases is 0.42, so that they can be compared fairly. Also, in the following analysis, as in~\cite{zakeri2024goal}, we call baseline the policy that follows the rule: If $e_t \geq \tau + \kappa$, then $a_t=3$, else $a_t=0$. This means the controller should sample and transmit if it has energy to it, or stay idle otherwise. The optimal policy is the one acquired through the RVI optimization. Finally, we compare the results of the \textit{Moving} monitor with the case where the controller has no knowledge about the monitor motion, and thus, utilizes a \textit{Constant} average RSR $q_s$ at all time slots.

\begin{figure}
    \centering
    \includegraphics[]{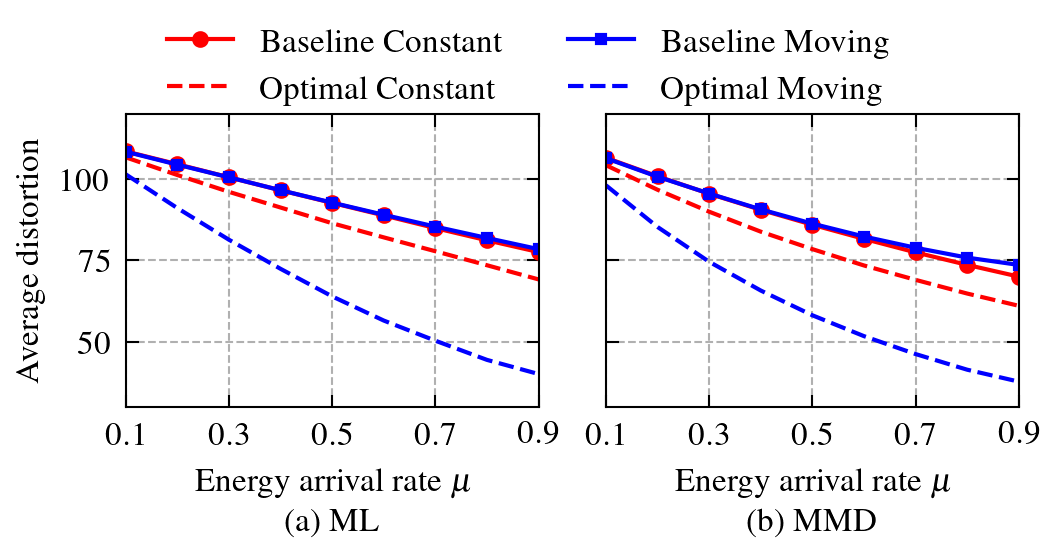}
    \caption{Average distortion with (a) ML and (b) MMD estimations as a function of the energy arrival rate $\mu$ for different policies and motion model from \eqref{eq:motion1}.}
    \label{fig:distortion_2states}
    \vspace{-0.5cm}
\end{figure}

\figurename~\ref{fig:distortion_2states} depicts the average distortion for ML and MMD estimation as a function of the energy arrival rate $\mu$ for the simplified motion monitor from~\eqref{eq:motion1}. First, we can see that the optimal policy in the moving scenario produces a smaller distortion compared to any of the other cases. We can attribute it to the policy being able to identify the best scenarios to transmit, where it produces distortion reduction going from 38\% to 42\% compared to the cases without channel knowledge. This is expected, as the motion matrix is simple, and with only two states, it can be straightforward to know that the controller should not transmit when the channel is at $q_s = 0$. However, the difference between both baseline schemes is minimal, although the moving scenario might produce cases with more failures in a row due to bad choices to transmit. Finally, similar to the results in~\cite{zakeri2024goal}, the MMD estimation outperforms the ML estimation, mostly because it uses knowledge from the source distortion to better estimate the states. However, because of the simplified motion monitor matrix, when there is more energy available in the optimal moving scenario, the monitor had to perform fewer estimations, and thus the estimation technique has a smaller impact. 

\begin{figure*}
    \centering
    \includegraphics[]{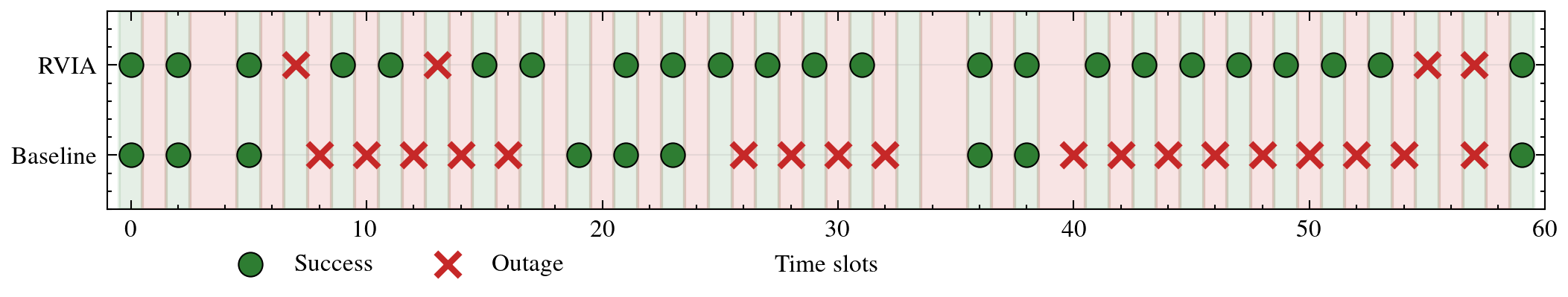}
    \caption{Simulation time window of 60 slots for the baseline and RVIA policies for $\mu=0.9$. Green and red background colors indicate the good and bad channel states, respectively, as defined in \eqref{eq:motion1}. Green circles denote successful transmissions, while red crosses indicate transmission outages.}
    \label{fig:success_2states}
    \vspace{-0.7cm}
\end{figure*}

\figurename~\ref{fig:success_2states} shows a time window of 60 slots from a simulation run using the baseline and RVIA policies for $\mu=0.9$. The energy arrivals and channel realizations are the same in both scenarios. Here becomes clear the advantages of the RVIA in choosing the correct moments to transmit, avoiding the bad states (red background). However, some outages still occur, since the success probability in the good state (green background) is 0.9, as seen in $q_s$ in \eqref{eq:motion1}. On the other hand, the baseline transmits whenever possible and will eventually transmit in the bad state.

\begin{figure}
    \centering
    \includegraphics[]{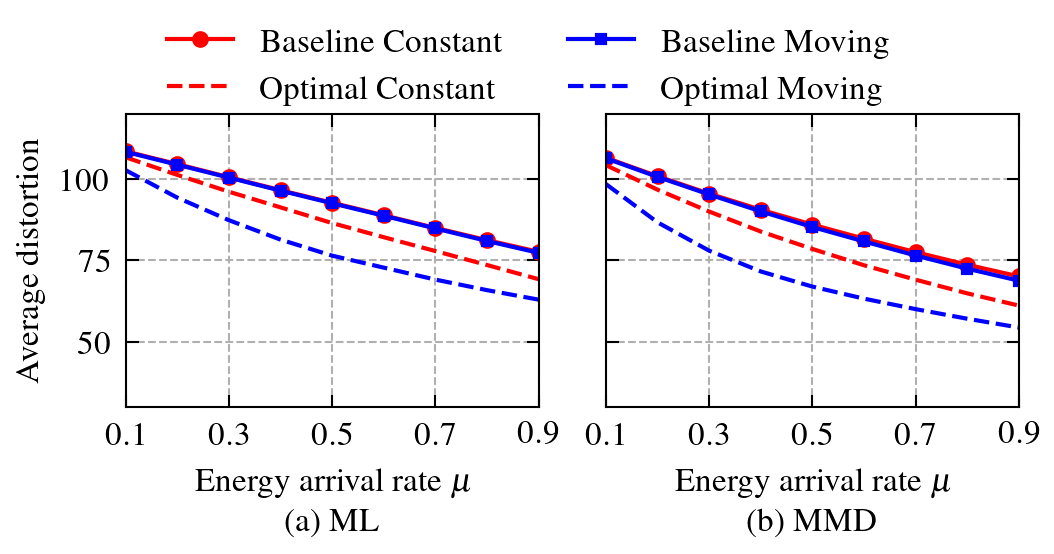}
    \caption{Average distortion with (a) ML and (b) MMD estimations as a function of the energy arrival rate $\mu$ for different policies and motion model from \eqref{eq:motion2}.}
    \label{fig:distortion_6states}
    \vspace{-0.5cm}
\end{figure}

\figurename~\ref{fig:distortion_6states} presents the same analysis as before, but for the motion monitor in~\eqref{eq:motion2}. First to notice is that optimal policy underperforms compared to the simplified monitor, despite both having the same average RSR. This degradation occurs because the monitor matrix is larger, causing the system to remain in suboptimal transmission states for longer. Thus, the controller has to decide to transmit in states other than the $q_s = 0.9$ case, unlike in the previous scenario. The baseline performs slightly better for the same reason: it now encounters more opportunities to transmit when $q_s \ne 0$. In this setting, incorporating channel knowledge (moving) yields distortion reductions between 9\% and 11\% compared to the case without such knowledge (constant).

\section{Conclusion} \label{sec:final_comments}

This work presented a goal-oriented remote monitoring framework for mobile receiver scenarios, extending previous stationary models to account for time-varying channels induced by receiver mobility. The system was formulated as a POMDP and solved via relative value iteration to obtain optimal sampling and transmission policies. Numerical results showed that exploiting receiver motion knowledge reduces average distortion by 10\% to 42\%, depending on the motion model complexity. These gains highlight the value of incorporating channel dynamics into policy optimization for energy-harvesting sensors in remote monitoring applications. Future directions include investigating imperfect channel state information and extending the framework to multi-sensor scenarios.

\section*{Acknowledgements}
This research was partially supported in Finland by the European Union through the Interreg Aurora project ENSURE-6G (Grant 20361812) and by the Research Council of Finland (RCF) through the projects 6G Flagship (Grant 369116), ECO-LITE (Grant 362782) and DYNAMICS (Grant 367702), in Brazil by CNPq (INCT STREAM 409179/2024-8, 305021/2021-4) and RNP/MCTI Brasil 6G (01245.020548/2021-07).
\vspace{-0.15cm}

\bibliographystyle{IEEEtran}
\bibliography{IEEEabrv,references}

@inproceedings{zakeri2024goal,
  title={Goal-oriented remote tracking of an unobservable multi-state {Markov} source},
  author={Zakeri, Abolfazl and Moltafet, Mohammad and Codreanu, Marian},
  booktitle={2024 IEEE Wireless Communications and Networking Conference (WCNC)},
  pages={1--6},
  year={2024},
  organization={IEEE}
}

@article{privault:2008:stochastic,
  title={Stochastic analysis of {Bernoulli} processes},
  author={Privault, Nicolas},
  year={2008}
}

@book{eliason1:993:Sage,
  title={Maximum likelihood estimation: Logic and practice},
  author={Eliason, Scott R},
  year={1993},
  publisher={Sage Publications}
}

@book{sigaud:Wiley:2013,
  title={{Markov} decision processes in artificial intelligence},
  author={Sigaud, Olivier and Buffet, Olivier},
  year={2013},
  publisher={John Wiley \& Sons}
}

@article{he2024non,
  title={Non-Terrestrial Network Technologies: Applications and Future Prospects},
  author={He, Peng and Lei, Hailong and Wu, Dapeng and Wang, Ruyan and Cui, Yaping and Zhu, Yan and Ying, Zhaopeng},
  journal={IEEE Internet of Things Journal},
  year={2024},
  publisher={IEEE}
}

@article{liu2021recent,
  title={Recent progress in the energy harvesting technology--from self-powered sensors to self-sustained {IoT}, and new applications},
  author={Liu, Long and Guo, Xinge and Liu, Weixin and Lee, Chengkuo},
  journal={Nanomaterials},
  volume={11},
  number={11},
  pages={2975},
  year={2021},
  publisher={MDPI}
}

@article{getu2024survey,
  title={A survey on goal-oriented semantic communication: Techniques, challenges, and future directions},
  author={Getu, Tilahun M and Kaddoum, Georges and Bennis, Mehdi},
  journal={IEEE Access},
  year={2024},
  publisher={IEEE}
}

@inproceedings{pappas:2021:ICAS,
  title={Goal-oriented communication for real-time tracking in autonomous systems},
  author={Pappas, Nikolaos and Kountouris, Marios},
  booktitle={2021 IEEE International Conference on Autonomous Systems (ICAS)},
  pages={1--5},
  year={2021},
  organization={IEEE}
}

@ARTICLE{zakeri:TWC:2024,
  author={Zakeri, Abolfazl and Moltafet, Mohammad and Leinonen, Markus and Codreanu, Marian},
  journal={IEEE Transactions on Wireless Communications}, 
  title={Minimizing the {AoI} in Resource-Constrained Multi-Source Relaying Systems: Dynamic and Learning-Based Scheduling}, 
  year={2024},
  volume={23},
  number={1},
  pages={450-466},
  doi={10.1109/TWC.2023.3278460}}

@ARTICLE{Zakeri:2025:TC,
  author={Zakeri, Abolfazl and Moltafet, Mohammad and Codreanu, Marian},
  journal={IEEE Transactions on Communications}, 
  title={Semantic-Aware Sampling and Transmission in Real-Time Tracking Systems: A {POMDP} Approach}, 
  year={2025},
  volume={73},
  number={7},
  pages={4898-4913},
  doi={10.1109/TCOMM.2024.3511935}}

@article{lutz:1996:Wiley,
  title={A {Markov} model for correlated land mobile satellite channels},
  author={Lutz, Erich},
  journal={International journal of satellite communications},
  volume={14},
  number={4},
  pages={333--339},
  year={1996},
  publisher={Wiley Online Library}
}

@article{aboderin:2015:modeling,
  title={Modeling land mobile satellite channel and mitigation of signal fading},
  author={Aboderin, Oluyomi and Alimi, Isiaka A},
  journal={American Journal of Mobile Systems, Applications and Services},
  volume={1},
  number={1},
  pages={46--53},
  year={2015}
}

@inproceedings{rougerie:2016:EuCAP,
  title={Mobile satellite propagation channels for {Ku} and {Ka} band},
  author={Rougerie, S and Lacoste, F and Montenegro-Villacieros, B},
  booktitle={2016 10th European Conference on Antennas and Propagation (EuCAP)},
  pages={1--5},
  year={2016},
  organization={IEEE}
}

@article{akinniyi:2017:Parallel,
  title={Modelling of Land Mobile Satellite Channel to Counter Channel Outage},
  author={Akinniyi, Akinade Olutayo and Tola, Awofolaju Tolulope and Olatunde, Oladepo},
  journal={Int. J. Distrib. Parallel Syst},
  volume={8},
  pages={1--21},
  year={2017}
}

@article{lee:2019:Access,
  title={Performance evaluation of high-frequency mobile satellite communications},
  author={Lee, Yonghwa and Choi, Jihwan P},
  journal={IEEE Access},
  volume={7},
  pages={49077--49087},
  year={2019},
  publisher={IEEE}
}

@article{fontan:2001:statistical,
  title={Statistical modeling of the {LMS} channel},
  author={Fontan, F Perez and V{\'a}zquez-Castro, Maryan and Cabado, C Enjamio and Garcia, J Pita and Kubista, Erwin},
  journal={IEEE Transactions on vehicular technology},
  volume={50},
  number={6},
  pages={1549--1567},
  year={2001},
  publisher={IEEE}
}

@inproceedings{hui:2008:new,
  title={A new five-state {Markov} model for land mobile satellite channels},
  author={Hui, M and Shen, DY and Cui, YN and others},
  booktitle={8th International Symposium on Antennas, Propagation and EM Theory, Kunming, China. Piscataway USA: IEEE Press},
  pages={1512--1515},
  year={2008}
}

@Article{Ijemaru:electronics:2021,
AUTHOR = {Ijemaru, Gerald K. and Ang, Kenneth L.-M. and Seng, Jasmine K. P.},
TITLE = {Mobile Collectors for Opportunistic Internet of Things in Smart City Environment with Wireless Power Transfer},
JOURNAL = {Electronics},
VOLUME = {10},
YEAR = {2021},
NUMBER = {6},
ARTICLE-NUMBER = {697},
ISSN = {2079-9292},
DOI = {10.3390/electronics10060697}
}

@ARTICLE{Wei:IoTJ:2022,
  author={Wei, Zhiqing and Zhu, Mingyue and Zhang, Ning and Wang, Lin and Zou, Yingying and Meng, Zeyang and Wu, Huici and Feng, Zhiyong},
  journal={IEEE Internet of Things Journal}, 
  title={{UAV}-Assisted Data Collection for Internet of Things: A Survey}, 
  year={2022},
  volume={9},
  number={17},
  pages={15460-15483},
  doi={10.1109/JIOT.2022.3176903}}

@ARTICLE{Wei:IoTJ:2021,
  author={Wei, Te and Feng, Wei and Chen, Yunfei and Wang, Cheng-Xiang and Ge, Ning and Lu, Jianhua},
  journal={IEEE Internet of Things Journal}, 
  title={Hybrid Satellite-Terrestrial Communication Networks for the Maritime Internet of Things: Key Technologies, Opportunities, and Challenges}, 
  year={2021},
  volume={8},
  number={11},
  pages={8910-8934},
  doi={10.1109/JIOT.2021.3056091}}

\end{document}